\documentclass[printer]{aa}

\usepackage{txfonts}
\usepackage[dvips]{graphicx}
\usepackage{natbib}

\begin{document}

\title{Self-consistent theory of turbulent transport in the solar tachocline}
\subtitle{II. Tachocline confinement}

\author{Nicolas Leprovost \and Eun-jin Kim}

\institute{Department of Applied Mathematics, University of Sheffield, Sheffield S3 7RH, UK}

\date{Received / Accepted }

\abstract{}{We provide a consistent theory of the tachocline confinement (or anisotropic momentum transport) within an hydrodynamical turbulence model. The goal is to explain helioseismological data, which show that the solar tachocline thickness is at most 5\% of the solar radius, despite the fact that, due to radiative spreading, this transition layer should have thickened to a much more significant value during the sun's evolution.}{Starting from the first principle with the  physically plausible assumption that turbulence is driven externally (e.g. by plumes penetrating from the convection zone), we derive turbulent (eddy) viscosity in the radial (vertical) and azimuthal (horizontal) directions by incorporating the crucial effects of shearing due to radial and latitudinal differential rotations in the tachocline.}{We show that the simultaneous presence of both shears induces effectively a much more efficient momentum transport in the horizontal plane than in the radial direction. In particular, in the case of strong radial turbulence (driven by overshooting plumes from the convection zone),
the ratio of the radial to horizontal eddy viscosity is proportional to ${\cal A}^{-1/3}$, where ${\cal A}$ is the strength of the shear due to radial differential rotation. In comparison, in the case of horizontally driven turbulence, this ratio becomes of order  $-\epsilon^2$, with negative radial eddy viscosity. Here, $\epsilon$ ($\ll 1$) is the ratio of the radial to latitudinal shear. The resulting anisotropy in momentum transport could thus be sufficiently strong to operate as a mechanism for the tachocline confinement against spreading.}{}

\keywords{Turbulence -- Sun: interior -- Sun: rotation}

\titlerunning{Tachocline confinement}

\maketitle

\section{Introduction}
One of the outstanding problems in solar physics is to understand the dynamics of the tachocline, a thin layer where the transition from latitudinal rotation in the convective zone to uniform rotation in the radiative interior takes place. This is particularly true since the tachocline links two regions of very different transport properties, thereby playing a crucial role in the overall angular momentum transport and chemical mixing on the course of the solar evolution. The tachocline is also crucial for the solar dynamo since the strong shear in the tachocline is believed to take part in the process by which the magnetic field of the sun is created. Helioseismic data \citep{Charbonneau99} constrain the thickness of this layer to be only a few percent of the solar radius. This poses a challenging problem since a radially localized shear flow (associated with the radial differential rotation) naturally tends to spread during the course of solar evolution, as shown by \citet{Spiegel92}, and should have reached a thickness much broader than what is observed today.

Some physical processes have been proposed to be responsible for the tachocline confinement. First, \citet{Spiegel92} have shown that if the turbulence is highly anisotropic (being much more vigorous in the horizontal direction than in the radial one), the spreading of the solar tachocline could be limited. More precisely, it is the turbulent transport of angular momentum which has to be anisotropic, meaning that the horizontal turbulent viscosity has to be much larger than the radial one. Physically, this would enable the differential rotation to be smoothed out by transport in the horizontal direction rather than in the vertical one. An alternative mechanism for the stabilization of the tachocline is due to magnetic field \citep{Rudiger97,Gough98,MacGregor99}: even a rather weak poloidal magnetic field (of the order $10^{-4} \, \mbox{G}$) in the solar interior could prevent the downward flow from the convection zone (driven, for example, by rotation-induced meridional circulation) from penetrating deeply into the radiative interior, thereby confining the differential rotation to a thin region of space {\bf and rendering a uniformly rotating interior}.

The relevance of these models (and any other that can be constructed with a physical process) to the sun is however not well asserted. In particular, the success of these models depends crucially on the values of transport coefficients such as turbulent viscosities and magnetic diffusivities, which are often crudely parameterized, unless their molecular values are used. Furthermore, the Gough \& McIntyre model is far to complex to be analyzed thoroughly and only a simplified version of it is actually tractable \citep{Garaud03}. In the Spiegel \& Zahn scenario, it is of prime importance to identify physical mechanisms which can lead to anisotropic momentum transport and then to derive eddy viscosities from first principles (or from Navier-Stokes equation). Note that while Spiegel \& Zahn invoked a strong stable stratification as a source of anisotropic turbulence, they did not derive the values of eddy viscosities.

The purpose of this paper is to provide a consistent theory of the anisotropic momentum transport. The physical mechanism that we invoke is the turbulence regulation by shearing effect, the so-called shear stabilization \citep{Burrell97,Hahm02,Kim04a}. It is basically because a shear flow acting on a turbulent eddy creates small scales in the direction orthogonal to the flow and thus enhances dissipation, reducing the transport and turbulence intensity \citep{Kim01,Kim03,Kim04b,Kim06}. Indeed, \cite{Kim05} has shown that a shear due to a stable radial differential rotation in the tachocline not only suppresses turbulent transport and turbulent intensity, but also leads to an \lq\lq effectively\rq\rq  \, anisotropic transport of particles via stronger reduction in radial transport compared to horizontal one. However, the key question of anisotropic momentum transport, necessary for the tachocline confinement, was not addressed in \citet{Kim05} since the effect of latitudinal differential rotation in the tachocline was simply neglected for simplicity. In this paper, we provide a theory of anisotropic momentum transport in the tachocline within an hydrodynamical turbulence model, by taking into account the crucial effects of both radial and latitudinal differential rotations in the tachocline. Specifically, we compute the transport properties of turbulence self-consistently under the physically plausible assumption that turbulence arises by an external forcing (e.g. from plumes penetrating from the convection zone). Most results will be derived in the limit of strong shear, but with latitudinal shear weaker than radial shear, as relevant to the tachocline.
As in \citet{Kim05}, we shall neglect the effects of stratification, magnetic field and rotation to elucidate the crucial role of shearing due to radial and latitudinal differential rotations in momentum transport. In particular, we shall show that shear alone can lead to anisotropic momentum transport, with a more efficient horizontal transport. While stratification, magnetic field and rotation can also contribute to anisotropic turbulence, these are outside the scope of this study and will be addressed in future papers.

The remainder of the paper is organized as follows: we provide our model of the tachocline and solve the governing equations in section \ref{Model}, then study the turbulence amplitude and eddy viscosities in section \ref{Amplitude} and section \ref{Viscosities} respectively. Section \ref{Conclusion} is devoted to our conclusions and discussions on the implications for the solar tachocline dynamics. Appendices contain some details of the algebra involved in solving our model and calculating the turbulent amplitude and viscosity.

\section{Model}
\label{Model}
To elucidate the effect of radial and latitudinal differential rotation on turbulent transport, we adopt a simplified model for the tachocline as in \citet{Kim05} and use a local Cartesian reference frame where $x$, $y$ and $z$ denote local radial, azimuthal, and latitudinal directions, respectively. We capture the radial and latitudinal differential rotation by a large-scale velocity field given by ${\bf U_0}(x,z) = - (x \mathcal{A}_x + z \mathcal{A}_z) \hat{y}$. Here, $\mathcal{A}_x = \partial_x U_0 > 0$ and $\mathcal{A}_z = \partial_z U_0 > 0$ are the strength of the radial and latitudinal shear due to radial and latitudinal differential rotation, respectively. In the quasi-linear approximation \citep{Moffatt78}, the Navier-Stokes equation  for the small-scale fluctuating velocity can be written:
\begin{eqnarray}
\nonumber
\partial_t {\bf v} - (x \mathcal{A}_x + z \mathcal{A}_z) \partial_y {\bf v} - (\mathcal{A}_x v_x + \mathcal{A}_z v_z) \hat{y} &=& - {\bf \nabla} p + \nu \nabla^2 {\bf v} + {\bf f} \; , \\ 
{\bf \nabla} \cdot {\bf v} &=& 0 \; .
\end{eqnarray}
Here ${\bf f}$ is the forcing at small scale and $\nu$ is the viscosity of the fluid. We will use a Fourier-transform with time-dependent wave number to account for the effect of shearing on eddies non-perturbatively [e.g. \cite{Kim05}]:
\begin{equation}
{\bf v}({\bf x},t) = \int d^3 {\bf k}(t) \;  e^{i {\bf k}(t) \cdot {\bf x}} \, \tilde{{\bf v}}({\bf k}(t),t) \; .
\end{equation}
Here, both radial and latitudinal wave-numbers evolve in time as follows:
\begin{equation}
k_x(t) = k_x(0) + \mathcal{A}_x t \qquad \mathrm{and} \qquad  k_z(t) = k_z(0) + \mathcal{A}_z t \; .
\end{equation}
To absorb the viscosity term, we set $\hat{\bf V} = \tilde{\bf V} \exp[\nu Q(t)]$  where $Q(t) = [k_x^3 / 3 k_y \mathcal{A}_x + k_y^2 t + k_z^3 / 3 k_y \mathcal{A}_z]$. The quasi-linear equations for the fluctuating part of the velocity field then become:
\begin{eqnarray}
\label{System1}
\partial_t \hat{v_x} &=& -i k_x \hat{p} + \hat{f_x} \; ,\\ \nonumber
\partial_t \hat{v_y} - (\mathcal{A}_x \hat{v_x} + \mathcal{A}_z \hat{v_z}) &=& -i k_y \hat{p} + \hat{f_y}  \; , \\ \nonumber
\partial_t \hat{v_z} &=& -i k_z \hat{p} + \hat{f_z} \; , \\ \nonumber
0 &=& k_x \hat{v_x} + k_y \hat{v_y} + k_z \hat{v_z}   \; .
\end{eqnarray}
To solve equation (\ref{System1}), we change the time variable from $t$ to $\tau = k_x(t) / k_y$ and introduce new variables: $\mathcal{A} =\mathcal{A}_x$, $\epsilon =\mathcal{A}_z / A_x$ and $\phi = [k_z(0) - \epsilon k_x(0)]/k_y$. For parameter values typical of the solar interior, we have: $\mathcal{A} = \Delta U_0 / \Delta X  \sim \Delta \Omega / (h/R_{\sun} ) = 3 \times 10^{-6} \, \mbox{s}^{-1}$ for the tachocline of thickness $4\%$ of the solar radius, and $\mathcal{A}_z = \Delta U / \Delta z \sim 4 \Delta \Omega \sim 8 \times 10^{-8} \, \mbox{s}^{-1}$. With these values, the ratio of the azimuthal and radial shear is a small parameter $\epsilon \sim 2.7 \times 10^{-2}$. We shall thus calculate the turbulent amplitude (section \ref{Amplitude}) and eddy viscosities (section \ref{Viscosities}) up to order $\epsilon^2$. In terms of the new variables, Eq. (\ref{System1}) becomes:
\begin{eqnarray}
\label{System2}
\mathcal{A} \partial_\tau \hat{v_x} &=& -i k_y \tau \hat{p} + \hat{f_x} \; ,\\ \nonumber
\mathcal{A} \partial_\tau \hat{v_y} - \mathcal{A} (\hat{v_x} + \epsilon \hat{v_z}) &=& -i k_y \hat{p} + \hat{f_y}  \; , \\ \nonumber
\mathcal{A} \partial_\tau \hat{v_z} &=& -i k_y (\epsilon \tau + \phi) \hat{p} + \hat{f_z} \; , \\ \nonumber
0 &=& \tau \hat{v_x} + \hat{v_y} + (\epsilon \tau + \phi) \hat{v_z}   \; .
\end{eqnarray}
The solution of this system of equations can be found after a long but straightforward algebra with the result (see appendix \ref{SolutionSyst} for details):
\begin{eqnarray}
\nonumber
\hat v_x &=& \int_{\tau_0}^{\tau}  d\tau_1 \; \frac{\hat{h_1}(\tau_1)}{\mathcal{A}} \Bigl\{ \frac{\gamma+\epsilon\phi \tau}{(\gamma+\epsilon^2) R(\tau)} + \frac{\epsilon \phi}{(\gamma+\epsilon^2)^{3/2}} \Bigl[T(\tau) -  T(\tau_1) \Bigr] \Bigr\} \\ 
&& - \int_{\tau_0}^{\tau}  d\tau_1 \frac{\hat{h_2}(\tau_1)}{\mathcal{A}} \frac{\epsilon}{\gamma+\epsilon^2} \; , \\ \nonumber
\hat v_y &=& \int_{\tau_0}^{\tau}  d\tau_1 \; \frac{\hat{h_1}(\tau_1)}{\mathcal{A}} \Bigl\{ \frac{\phi^2}{(\gamma+\epsilon^2)^{3/2}}\bigl(T(\tau) - T(\tau_1)\bigr) \\ \nonumber
&& - \frac{(\epsilon^2+1)\tau + \epsilon \phi}{(\gamma+\epsilon^2)R(\tau)} \Bigr\} - \int_{\tau_0}^{\tau}  d\tau_1 \frac{\hat{h_2}(\tau_1)}{\mathcal{A}} \frac{\phi}{\gamma+\epsilon^2} \; , \\ \nonumber
\hat v_z &=& \int_{\tau_0}^{\tau}  d\tau_1 \; \frac{\hat{h_1}(\tau_1)}{\mathcal{A}} \Bigl\{ \frac{\epsilon - \phi \tau}{(\gamma+\epsilon^2) R(\tau)} - \frac{\phi}{(\gamma+\epsilon^2)^{3/2}} \Bigl[T(\tau)-  T(\tau_1) \Bigr] \Bigr\} \\ \nonumber 
&& + \int_{\tau_0}^{\tau}  d\tau_1 \frac{\hat{h_2}(\tau_1)}{\mathcal{A}} \frac{1}{\gamma+\epsilon^2}  \; .
\end{eqnarray}
{\bf Here, 
\begin{eqnarray}
R(\tau) &=& (\epsilon^2+1) \tau^2 + 2 \epsilon \phi \tau + \gamma \; , \qquad  \gamma = 1 + \phi^2 \; , \\ \nonumber 
T(\tau) &=& \arctan\Bigl(\frac{(\epsilon^2+1) \tau + \epsilon \phi}{\sqrt{\gamma+\epsilon^2}}\Bigr) \; , \\ \nonumber 
\hat{h_1}(\tau) &=& [\gamma + \phi \epsilon \tau]\hat{f_x} - [(1+\epsilon^2) \tau + \phi \epsilon] \hat{f_y} + (\epsilon-\phi \tau) \hat{f_z} \; , \\ \nonumber 
\hat{h_2}(\tau) &=& - \epsilon \hat{f_x}(\tau) - \phi \hat{f_y}(\tau) + \hat{f_z}(\tau) \; .
\end{eqnarray}
}

\section{Turbulence amplitude}
\label{Amplitude}
We first examine how turbulence amplitude is affected by latitudinal shear, comparing the results with \cite{Kim05}. In order to compute $\langle v_i^2 \rangle$, we assume the forcing to be incompressible (thus all the quantities can be expressed in terms of the $x$ and $z$ components only) with the statistics that is spatially homogeneous and temporally short correlated with the correlation time $\tau_f$. Specifically, we take:
\begin{equation}
\langle \tilde{f}_i({\bf k_1},t_1) \tilde{f}_j({\bf k_2},t_2) \rangle = \tau_f \, (2\pi)^3 \delta({\bf k_1}+{\bf k_2}) \, \delta(t_1-t_2) \, \psi_{ij}({\bf k_2}) \; ,
\end{equation}
for $i$ and $j$ = $1,2$ or $3$. The $\psi_{ij}$ functions are the power spectrum of the forcing.

A typical solar value of the molecular viscosity is $\nu \sim 10^2 \, \mbox{cm}^2 \mbox{s}^{-1}$  and thus the parameter $\xi = \nu k_y^2 / \mathcal{A}$ is a small quantity provided that $k_y < 10^{-4} \, \mbox{cm} \sim 10^{-6} / H_0$ where $H_0$ is the pressure scale height at the bottom of the convection zone. The relevant length scale in the azimuthal direction being probably larger than $10^4 \mbox{cm}$, this permits us to consider the strong shear limit $\xi \ll 1$. In addition, since $\epsilon$ is a small parameter, as noted previously, we expand the result up to $\epsilon^2$, keeping only the dominant term in $\xi$ and $\epsilon$ for each component of the forcing $\psi_{ij}$. A long but straightforward algebra, very similar to that of the calculation of the turbulent viscosities (see next section and appendix \ref{Calcul2}), then gives us the turbulence amplitude in the strong shear limit $\xi \ll 1$ to leading order in $\xi$ and $\epsilon$:
\begin{eqnarray}
\label{TurbAmpli}
\langle v_x^2 \rangle &=& \frac{\tau_f}{(2 \pi)^3 \mathcal{A}} \int d^3 {\bf k}  \Bigl\{ \frac{g+a^2}{2 g} \bigl[-a + \frac{g+a^2}{\sqrt{g}} \kappa_0 \bigr] \psi_{11}({\bf k}) \\ \nonumber
&& \qquad + \frac{(g+a^2) \epsilon}{g} \bigl[-a + \frac{a^2-g}{\sqrt{g}} \kappa_0 \bigr] \psi_{13}({\bf k})  + \epsilon^2 \mathcal{G}_0 \, \psi_{33}({\bf k}) \Bigl\}  \; ,\\ \nonumber
\langle v_y^2 \rangle &=& \frac{\tau_f}{(2 \pi)^3 \mathcal{A}} \int d^3 {\bf k} \; \mathcal{G}_0 \; \Bigl\{ \frac{b^4}{g^3} \Bigl((g+a^2)^2 \kappa_0^2 + g a^2\Bigr) \, \psi_{11}({\bf k})  \\ \nonumber && \qquad + \frac{2 b^3 a}{g}  \psi_{13}({\bf k}) + b^2  \psi_{33}({\bf k})  \Bigr\} \; ,\\ \nonumber
\langle v_z^2 \rangle &=& \frac{\tau_f}{(2 \pi)^3 \mathcal{A}} \int d^3 {\bf k} \; \mathcal{G}_0 \; \Bigl\{ \frac{b^2}{g^3} \Bigl((g+a^2)^2 \kappa_0^2 + g a^2 \Bigr) \, \psi_{11}({\bf k}) \\ \nonumber 
&& \qquad + \frac{2 b a}{g}  \psi_{13}({\bf k}) +  \psi_{33}({\bf k})  \Bigr\} \; .
\end{eqnarray}
{\bf Here, 
\begin{eqnarray}
a&=&k_x/k_y \; , \qquad b=k_z/k_y \; , \qquad g=1+b^2 \; , \\ \nonumber
\kappa_0 &=& \pi/2 - \arctan(a/\sqrt{g}) = \arctan(\sqrt{g}/a) \; , \\ \nonumber
\mathcal{G}_0 &=& \Gamma(1/3)/3 (3/2\xi)^{1/3} \; , \qquad \xi = \nu k_y^2 / \mathcal{A} \; .
\end{eqnarray}
}
In Eq. (\ref{TurbAmpli}), we kept only the leading order contributions in $\xi$ and $\epsilon$ for the terms proportional to $\psi_{11}$, $\psi_{13}$ and $\psi_{33}$. For instance, for the component proportional to $\psi_{33}$ in the radial turbulence amplitude, we expanded the quantities up to order $O(\epsilon^2)$. It is worth noting that $\langle v_x^2 \rangle$ is always positive, as it should: while the second term proportional to $\psi_{13}$ is not always positive, in order for the sum of the two terms proportional to $\psi_{11}$ and $\psi_{13}$ to be negative, $\psi_{13}$ has to be at least $\epsilon^{-1}$ larger than $\psi_{11}$; in that case, $\psi_{33}$ would be at least of order $\epsilon^{-2}$ and thus the last term would dominate the other two terms, making $\langle v_x^2 \rangle$ positive (recall that $\xi \ll 1$ and thus $\mathcal{G}_0 \gg 1$). In comparison, the horizontal turbulent velocities in Eq. (\ref{TurbAmpli}) are obviously positive and are the same as that obtained by \cite{Kim05} in the case without a latitudinal shear.

Eq. (\ref{TurbAmpli}) indicates that the amplitude of the three components of the velocity depends not only on the radial and latitudinal shear (shearing rate $\mathcal{A}$ and $\epsilon \mathcal{A}$) but also on the typical wavenumber ${\bf k}$ and the power spectrum of the forcing $\psi_{ij}({\bf k})$. We first examine how various components are affected by the radial and latitudinal shear. As can easily been seen from Eq. (\ref{TurbAmpli}), the amplitude of all three components of the turbulent velocity is reduced due to radial shear, becoming
very small as $\mathcal{A}$ increases. Specifically, the horizontal turbulence is reduced by a factor $\mathcal{A} \mathcal{G}_0^{-1} \propto \mathcal{A}^{2/3}$, the inclusion of the latitudinal shear having no effect on the horizontal turbulence amplitude. In comparison, the radial turbulence velocity has a small contribution of order $O(\epsilon^2)$ from $\psi_{33}$ because of a latitudinal shear. In order to understand this, it is instructive to consider a new reference frame such that the shear flow
${\bf U}_0 {\hat y}$, depending on both $x$ and $z$, becomes a function of only one coordinate $x'$. This coordinate transformation can easily be obtained by rotating $x$ and $z$ axis by the angle $\theta = \arctan(\mathcal{A}_z/\mathcal{A}_x) \sim \epsilon$ around the $y$ axis (see figure \ref{ChgtRef}) as follows: 
\begin{figure}[t]
\resizebox{\hsize}{!}{\includegraphics{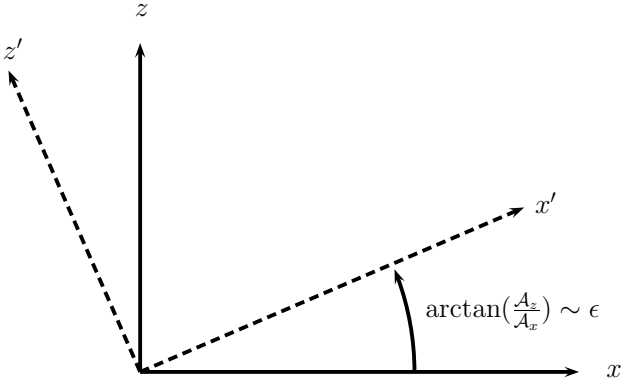}}
\caption{\label{ChgtRef} Sketch of the coordinate transformation from $(x,z)$ to $(x',z')$. In the new referential, the shear is purely radial (see main text for details).}
\end{figure}
\begin{equation}
\left(
\begin{array}{c}
x' \\
z'
\end{array} \right)
=
\left(
\begin{array}{c c}
\cos\theta  & \sin\theta \\
-\sin\theta & \cos\theta
\end{array} \right)
\left(
\begin{array}{c}
x \\
z
\end{array} \right)
\sim
\left(
\begin{array}{c c}
1  & \epsilon \\
- \epsilon & 1
\end{array} \right)
\left(
\begin{array}{c}
x \\
z
\label{transf}
\end{array} \right) \; ,
\end{equation}
where we used $\epsilon \ll 1$. In the new $(x',z')$ coordinates, the large-scale velocity field is given by ${\bf U_0'} = - (\mathcal{A}_x^2 + \mathcal{A}_z^2)^{1/2} x' \hat{y'}$, depending only on the radial direction, thus becoming a purely radial shear. Thus, in the $(x',z')$ coordinate, we can use the results of \cite{Kim05} that $\langle v_z'^2 \rangle  \propto \mathcal{A}^{-1} \mathcal{G}_0 \sim \mathcal{A}^{-2/3}$ and $\langle v_x'^2 \rangle \propto \mathcal{A}^{-1}$. By relating these to $\langle v_x^2 \rangle$ and $\langle v_z^2 \rangle$ via Eq. (\ref{transf}), we easily obtain the term proportional to $\psi_{33}$ in $\langle v_x^2 \rangle$ in Eq. (\ref{TurbAmpli}).  That is, the latter is the result of the dependence of $\langle v_z'^2 \rangle$ in the $(x',z')$ reference frame with an additional factor of $\epsilon^2 \sim \sin^2\theta$ coming from the projection of this term onto the $x$ axis. We also note that $\langle v_z'^2 \rangle = \langle v_z^2 \rangle$ for $\epsilon \ll 1$, consistent with Eq. (\ref{TurbAmpli}). We will further use this geometrical argument when discussing the turbulent viscosities.

Keeping these in mind, we now discuss the influence of the different types of forcing on the turbulent intensity. If the turbulent motions are primarily driven in the radial direction ($\psi_{33} = \psi_{13} = 0$), the latitudinal shear has no effect, and the results are the same as in the case of a radial shear only: the radial turbulence is reduced by a factor $\mathcal{A}$ and thus, the ratio of the radial to horizontal turbulence amplitude (taking the $y$ direction as well as the $z$ direction) is of order $\mathcal{A}^{-1/3}$. Alternatively, if the small-scale flow is mainly driven in the horizontal direction ($\psi_{11} = \psi_{13} = 0$), this component $\langle v_x^2 \rangle $ is reduced by a factor of  $\epsilon^2 \mathcal{A}^{-2/3}$, with the ratio $\langle v_x^2 \rangle /\langle v_z^2 \rangle  \propto \epsilon^{2}$, for the reason that
was just discussed. Again, note that this is different from the case without latitudinal shear where the radial turbulence intensity was null (at the order of the approximation). The ratio $\langle v_x^2 \rangle /\langle v_z^2 \rangle  \propto \epsilon ^2$ is still small in the case of the sun but interestingly depends only on the ratio of the strength of the azimuthal and radial shear but not on their absolute magnitudes.

\section{Turbulent viscosity}
\label{Viscosities}
We now proceed to the calculation of the radial $\nu_T^{xx}$ and horizontal viscosity $\nu_T^{zz}$ defined as: $\langle v_x v_y \rangle = - \nu_T^{xx} \partial_x U_0 = \nu_T^{xx} \mathcal{A}$ and $\langle v_z v_y \rangle = - \nu_T^{zz} \partial_z U_0 = \nu_T^{zz} \epsilon \mathcal{A}$. From the previous section, one might naively estimate the ratio of these two to be of order:
\begin{equation}
\frac{\langle v_x v_y \rangle}{\langle v_y v_z \rangle} \sim \sqrt{\frac{\langle v_x^2 \rangle}{\langle v_z^2 \rangle}} \sim  \xi^{1/6} \quad \mathrm{or} \quad \epsilon \; ,
\label{ratio}
\end{equation}
for a $x$-dominant and $z$-dominant forcing respectively. However, the momentum fluxes $\langle v_x v_y \rangle$ and $\langle v_y v_z \rangle$ depend not only on the velocity amplitudes, but also how closely different components of the velocity are correlated (i.e. the phase-relation). That is, the ratio of the two should be given as follows:
\begin{equation}
\label{ratio1}
\frac{\langle v_x v_y \rangle}{\langle v_y v_z \rangle}  = \sqrt{\frac{\langle v_x^2 \rangle}{\langle v_z^2 \rangle}} \frac{\cos{\delta_{xy}}}{\cos{\delta_{zy}}} \; ,
\end{equation}
where $\cos{\delta_{xy}}$ and $\cos{\delta_{zy}}$ (the so-called cross phase) represent the phase-relation between $x$ and $y$ and $z$ and $y$ components of the velocity, respectively. Thus, the estimate given in Eq. (\ref{ratio}) may not be true
if the cross-phase $\cos{\delta_{xy}}$ and $\cos{\delta_{zy}}$ are affected in an anisotropic way due to the different strength in the radial and latitudinal shear. We shall show that this is indeed the case. Specifically, $\cos{\delta_{xy}}/\cos{\delta_{zy}} \propto \xi^{1/6}$ when the turbulence is driven mainly in the $x$-direction while $\cos{\delta_{xy}}/\cos{\delta_{zy}} = -1$ when driven in the horizontal direction.

After a long, but straightforward algebra, we can obtain eddy viscosities
in the following form (see appendix \ref{Calcul2} for details):
\begin{eqnarray}
\label{EddyViscos}
\nonumber
\nu_T^{xx} &=&  \frac{\tau_f}{(2 \pi)^3 \mathcal{A}^2} \int d^3 {\bf k}  \Bigl\{ \frac{g+a^2}{2g^2} \bigl[b^2(g^2+a^2)\kappa_0^2 - g \\ \nonumber 
&& \qquad \qquad \qquad - \frac{4 \epsilon b^3 a}{\sqrt{g}} \kappa_0  \mathcal{G}_0 \bigr] \psi_{11}({\bf k}) \\ \nonumber
&& - \frac{(g+a^2) 2 \epsilon b^2}{g^{3/2}} \kappa_0 \mathcal{G}_0 \psi_{13}({\bf k}) - \frac{2 \epsilon^2 b^2 (g+a^2)}{g^{3/2}} \kappa_0 \mathcal{G}_0 \psi_{33}({\bf k}) \Bigr\} \; ,\\ 
\nu_T^{zz} &=& \frac{\tau_f}{(2 \pi)^3 \mathcal{A}^2} \int d^3 {\bf k} \frac{1}{\epsilon} \Bigl\{ \frac{g+a^2}{2g^2} \bigl[b a + \frac{4 \epsilon b^4}{\sqrt{g}} \kappa_0  \mathcal{G}_0 \bigr] \psi_{11}({\bf k}) \\ \nonumber 
&& \qquad + \frac{2 b^2 (g+a^2)}{g^{3/2}} \kappa_0 \mathcal{G}_0 \Bigl(\psi_{13}({\bf k}) + \epsilon \psi_{33}({\bf k})\Bigr) \Bigr\} \; .
\nonumber
\end{eqnarray}
Here, again, $a=k_x/k_y$, $b=k_z/k_y$, $g=1+b^2$, $\kappa_0 = \pi/2 - \arctan(a/\sqrt{g}) = \arctan(\sqrt{g}/a)$, $\mathcal{G}_0 = \Gamma(1/3)/3 (3/2\xi)^{1/3}$ and $\xi = D k_y^2 / \mathcal{A}$.

\begin{figure}[t]
\resizebox{\hsize}{!}{\includegraphics{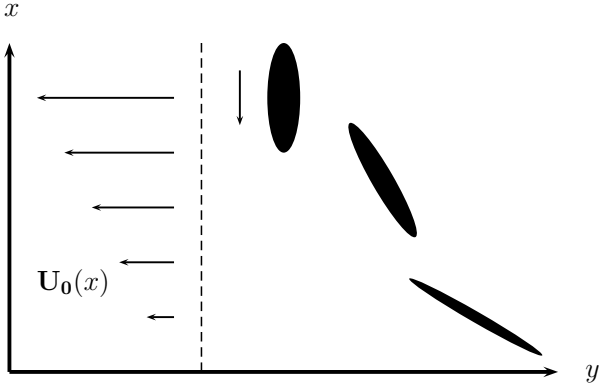}}
\caption{\label{Plumes} Effect of the large-scale shear (left) on plumes coming from the convection zone (right). In addition to the tilting induced by the large-scale shear, the turbulent velocity field induces stochastic distortion of the eddies which is not drawn here for simplicity.}
\end{figure}

Eq. (\ref{EddyViscos}) clearly shows that both $\nu_T^{xx}$ and $\nu_T^{zz}$ are reduced by shearing, becoming very small as $\mathcal{A}$ increases. The exact scalings of the eddy viscosities however depend on the properties of the forcing since $\mathcal{G}_0
\propto \xi^{-1/3} \propto \mathcal{A}^{1/3}$ appears in some of the coefficients proportional to $\psi_{11}$ and $\psi_{13}$ in Eq. (\ref{EddyViscos}) with both positive and negative signs, making the estimate rather complex in the case where the forcing is mainly driven radially by $\psi_{11}$. In order to obtain a transparent scaling relation in this case, it is illuminating to consider a physically plausible case where the turbulent plumes, coming from the convection zone, are highly elongated in the $x$-direction, with small-scale structures in the radial direction being mainly created by radial shear (see figure \ref{Plumes}). In this case, it is very likely that the forcing will select small wave numbers in the radial direction, permitting us to crudely set $a=k_x / k_y \sim 0$ in Eq. (\ref{EddyViscos}). Mathematically, this is justified if the support of the $\psi_{ij}$ functions is localized near the line $k_x/k_y =0$. In that case, the radial viscosity does not contain terms involving $\epsilon$ or $\xi$ while the horizontal component is proportional to $\mathcal{G}_0 = \Gamma(1/3)/3 (3/2\xi)^{1/3}$. Therefore, the ratio of these two viscosities is of order $\nu_T^{xx}/\nu_T^{zz} \propto \xi^{1/3} \ll 1$. In other words, the momentum transport becomes anisotropic with a much larger horizontal transport compared to the radial one.

Interestingly, the contribution from $\psi_{33}$ to $\nu_T^{xx}$ comes with the opposite sign to that from $\psi_{11}$ while the contribution from $\psi_{33}$ to $\nu_T^{zz}$ come with the same sign as that from $\psi_{11}$. Thus, as $\psi_{33}$ takes a non-vanishing value due to horizontal forcing, it will further decrease $\nu_T^{xx}$ while increasing $\nu_T^{zz}$ until $\nu_T^{xx}$ vanishes and then becomes negative. In other words, the anisotropy in momentum transport becomes even stronger with $\nu_T^{xx}/\nu_T^{zz} \ll \xi^{1/3}$. In the extreme limit where the turbulence is solely driven horizontally with $\psi_{11} = \psi_{13} = 0$,
$\nu_T^{xx}/\nu_T^{zz} \propto -\epsilon^2$. Note that this is true in general, regardless of the form of power spectrum of $\psi_{33}$.  Since $\epsilon^2 \ll 1$, the ratio of the magnitude of the two
is small. It is important to note that the two viscosities have different signs, with
the radial (vertical) viscosity being negative while the horizontal one is positive. 
To understand these results, it is useful to consider the coordinate $(x',z')$ given in Eq. (\ref{transf}) where the shear flow depends only on $x'$. By an elementary calculation, we can easily show that the turbulent viscosity $\nu_T^{x'x'}$ in the $(x',z')$ coordinate can be expressed in terms of $\nu_T^{xx}$ and $\nu_T^{zz}$ in the $(x,z)$ coordinate as follows:
\begin{equation}
\label{ViscoBis}
\nu_T^{x'x'} = \frac{1}{\mathcal{A}_x^2 + \mathcal{A}_z^2}\left[\mathcal{A}_x^2 \nu_T^{xx} + \mathcal{A}_z^2 \nu_T^{zz}\right] \; .
\end{equation}
We now recall that in the $(x',z')$ coordinate with a purely radial shear, $\nu_T^{x'x'} = 0$ in the case of a purely horizontal forcing ($\psi_{11} = \psi_{13} = 0$), as shown by \cite{Kim05}. Thus, requiring that the right hand side of equation (\ref{ViscoBis}) be zero immediately gives us the result $ \nu_T^{xx}/\nu_T^{zz} \propto - \epsilon^2$, as found in this paper. This is an interesting result since the negative viscosity here  is not related to the 2-dimensionality of the flow in any sense, being merely due to the fact that a special rotation of the Cartesian frame maps our problem to the case of a purely radial shear. 

To summarize, for a reasonable choice of the forcing, $\nu_T^{xx}/\nu_T^{zz} \propto \xi^{1/3}\ll 1$ and $-\epsilon^2$ for radial and horizontal forcing, respectively. In these two limiting cases, the ratio of the cross-phase given in Eq.\ (\ref{ratio1}) satisfies $\cos{\delta_{xy}}/\cos{\delta_{zy}} \propto \xi^{1/6}$ and $ -1$, respectively. It is worth noting that the small value of $\cos{\delta_{xy}}/\cos{\delta_{zy}}\propto \xi^{1/6}$ in the former case physically makes sense since a stronger decorrelation between velocity components in the radial direction and horizontal plane is expected because of a stronger radial shear. These results thus suggest that horizontal momentum transport can be much more efficient than radial transport due to radial and latitudinal differential rotations.


\section{Conclusions and discussion}
\label{Conclusion}
We presented a novel mechanism for the anisotropic angular momentum transport to explain tachocline confinement. Starting from the first principle, we have consistently derived eddy viscosities and turbulence amplitude by incorporating the crucial effects of both radial and latitudinal differential rotations. We have shown that even in the absence of stratification and magnetic field, the shear induced by differential rotation alone can lead to anisotropic momentum transport. Specifically, when the turbulence is mainly driven radially by overshooting plumes coming from the convection zone, the ratio of the radial to horizontal eddy viscosities can be proportional to $\xi^{1/3}$, becoming very small as the radial shear $\mathcal{A}$ increases. Here, $\xi = \nu k_y^2 / \mathcal{A}$  and $\mathcal{A}$ is the shearing rate in the radial direction. In comparison, in the case where the turbulence is mainly driven horizontally, this ratio is proportional to $-\epsilon^2$ where $\epsilon \; (\ll 1)$ is the ratio of the shearing rate in the latitudinal direction to that in the radial one. In this case, the radial eddy viscosity becomes negative. These results thus suggest that horizontal momentum transport can be much more efficient than radial transport, with a strong anisotropic momentum transport, as a result of the shearing effects due to radial and latitudinal differential rotations.

As mentioned in introduction, the anisotropic momentum transport has important implications for the dynamics of the tachocline. \citet{Spiegel92} have shown that the radiative spreading of the tachocline could be limited if the ratio of the radial to horizontal diffusivities satisfies the following relation:
\begin{equation}
\frac{\nu_T^{xx}}{\nu_T^{zz}} \ll \Bigl(\frac{h}{r_0}\Bigr)^2 \sim 3 \times 10^{-3} \; ,
\end{equation}
where $r_0$ is the radius at which the tachocline is located and $h$ its thickness. The estimate on the right hand side has been obtained using helioseismologic values of \cite{Charbonneau99}. Indeed, this criterion can easily be satisfied for a reasonable choice of the forcing as follows. First, in the case of radial forcing which is dominated by the components with $k_x/k_y \ll 1$, the Spiegel \& Zahn criterion is easily met for a characteristic length-scale in the azimuthal direction $L_y = 2 \pi k_y^{-1} \gg 2 \pi (r_0/h)^{3} (\nu / \mathcal{A})^{1/2} \sim 3 \times 10^5 \, \mbox{m}$ for typical values of the solar tachocline $\mathcal{A} \sim 3 \times 10^{-6} \, \mbox{s}^{-1}$.  In the case of horizontal forcing, the ratio of radial to horizontal eddy viscosities becomes of order $\epsilon^2 \sim 7 \times 10^{-4}$. Furthermore, in this case, the radial eddy viscosity becomes negative, reinforcing the presence of radial shear with sharp gradient in a localized region. Therefore, these results suggest that the anisotropy in momentum transport could be strong enough to operate as a mechanism for the tachocline confinement against spreading. {\bf Our results, however, cannot offer a mechanism for a uniform rotation in the solar interior. As noted in the introduction, this could be due to the presence of magnetic fields in this region  \citep{Rudiger97,Gough98,MacGregor99}.}

We note that in this paper we have neglected some effects which could be important in the
lower part of the tachocline. For example, the presence of stratification or rotation can generate a wave-like turbulence and consequently severely affects its transport properties. The effect of Coriolis forces has been studied in the context of the transport of angular momentum in the convection zone to elucidate the permanence of a differential rotation profile in that region. It has been shown that, in the strong rotation limit, the turbulent viscosity is more reduced by a factor of 4 \citep{Kichatinov94} in the direction perpendicular to the rotation axis compared to the one parallel to the rotation axis. Furthermore, rotation induces a non-diffusive part in the turbulent viscosity (the so-called $\Lambda$ effect) which prevents the solid rotation to be a solution of the turbulent momentum equation. We can expect a density stratification to have a similar effect to the Coriolis force as the linear equations are the same \citep{Greenspan68} once the vertical and horizontal directions are interchanged. In that case, we expect the component in the direction of the stratification to be more reduced than in the plane of constant density. {\bf In fact, such an anisotropic eddy viscosity has been observed in a recent numerical simulation of a stably stratified turbulence driven by penetrative convection with an imposed shear \citep{Miesch03}. In particular, the angular momentum transport was found to be diffusive in the latitudinal direction while anti-diffusive in the radial direction, with turbulence mixing potential vorticity [see also \citet{McIntyre03}].} Another important effect is due to magnetic fields in the tachocline (as suggested by solar-dynamo models) by imposing  Alfv\`en-like structure in the properties of turbulence. Furthermore, the simultaneous presence of toroidal magnetic field and latitudinal differential rotation can lead to a large-scale joint instability \citep{Gilman97,Dikpati04} which could further enforce an anisotropic turbulence. The interplay between shear and rotation is under current investigation \citep{Leprovost06e} and the related problems of stratification and magnetic field will be published in future papers.


\appendix
\section{Solution of system (\ref{System1})}
\label{SolutionSyst}
Using the new field $\hat{u} = \hat{v_x} + \epsilon \hat{v_z}$, we obtain the following equations:
\begin{eqnarray}
\label{Newform}
\mathcal{A} \, \partial_\tau [R(\tau) \hat{u}] &=& \hat{h_1} \; , \\ \nonumber
\mathcal{A} [\gamma + \phi \epsilon \tau] \partial_\tau \hat v_x &=& - \mathcal{A} \partial_\tau[\tau^2 \hat{u}] + [\gamma + \phi \epsilon \tau] \hat{f_x} - \tau \hat{f_y} - \phi \tau \hat{f_z} \; ,
\end{eqnarray}
where $\gamma = 1 + \phi^2$, $R(\tau) = (\epsilon^2+1) \tau^2 + 2 \epsilon \phi \tau + \gamma$ and $\hat{h_1} = [\gamma + \phi \epsilon \tau]\hat{f_x} - [(1+\epsilon^2) \tau + \phi \epsilon] \hat{f_y} + (\epsilon-\phi \tau) \hat{f_z}$. The first equation can be readily integrated to obtain $\hat{u} = \int d\tau_1 \; \hat{h_1}(\tau_1) / \mathcal{A} R(\tau_1)$. Then, the second equation can be used to obtain $\hat{v_x}$. Here, we provide some of the main steps in the calculation. First, we calculate the first term on the RHS of Eq. (\ref{Newform}):
\begin{equation}
\label{Intermed1}
\partial_\tau[\tau^2 \hat{u}] = \frac{\tau^2 \hat{h_1}(\tau)}{\mathcal{A} R(\tau)} + \frac{2 \tau [\gamma + \epsilon \phi \tau]}{R(\tau)^2} \int_{\tau_0}^\tau d\tau_1 \; \frac{\hat{h_1}(\tau_1)}{\mathcal{A}} \; .
\end{equation}
Then plugging Eq. (\ref{Intermed1}) in the second equation of (\ref{Newform}), we obtain:
\begin{eqnarray}
\hat v_x &=&  - \int_{\tau_0}^\tau d\tau_2 \; \frac{2 \tau_2}{R(\tau_2)^2} \int_{\tau_0}^{\tau_2}  d\tau_1 \; \frac{\hat{h_1}(\tau_1)}{\mathcal{A}}  \\ \nonumber 
&& + \int_{\tau_0}^\tau d\tau_1 \frac{[R(\tau_1)-\tau_1^2] \hat{f_x} - \tau_1 \hat{f_y} - \tau_1 (\phi+\epsilon \tau_1) \hat{f_y}}{\mathcal{A}  R(\tau_1)}  \; , \\ \nonumber
&=&  - \int_{\tau_0}^{\tau}  d\tau_1 \; \frac{\hat{h_1}(\tau_1)}{\mathcal{A}} \int_{\tau_1}^\tau d\tau_2 \; \frac{2 \tau_2}{R(\tau_2)^2}  \\ \nonumber 
&& + \int_{\tau_0}^\tau d\tau_1 \frac{[R(\tau_1)-\tau_1^2] \hat{f_x} - \tau_1 \hat{f_y} - \tau_1 (\phi+\epsilon\tau_1) \hat{f_y}}{\mathcal{A}  R(\tau_1)}  \; , \\ \nonumber
&=& \int_{\tau_0}^{\tau}  d\tau_1 \; \frac{\hat{h_1}(\tau_1)}{\mathcal{A}} \Bigl\{ \frac{\gamma+\epsilon\phi \tau}{(\gamma+\epsilon^2) R(\tau)} + \frac{\epsilon \phi}{(\gamma+\epsilon^2)^{3/2}} \Bigl[ T(\tau)-  T(\tau_1) \Bigr] \Bigr\} \\ \nonumber 
&& - \int_{\tau_0}^{\tau}  d\tau_1 \frac{\hat{h_2}(\tau_1)}{\mathcal{A}} \frac{\epsilon}{\gamma+\epsilon^2} \; .
\end{eqnarray}
Here we used the formula (2.175) and (2.172) of \citet{GR65} to calculate the integrals; $T(\tau) = \arctan([(\epsilon^2+1) \tau + \epsilon \phi] / \sqrt{\gamma+\epsilon^2})$ and $\hat{h_2}(\tau) = - \epsilon \hat{f_x}(\tau) - \phi \hat{f_y}(\tau) + \hat{f_z}(\tau)$. The two other components of the velocity can easily be obtained by using $\hat{v_z} = (\hat{u} - \hat{v_x})/\epsilon$ and $\hat{v_y} = - \tau \hat{v_x} - (\epsilon \tau + \phi) \hat{v_z}$.

\section{Calculation of the turbulent viscosities}
\label{Calcul2}
We symbolically write $\langle v_x v_y \rangle$ and $\langle v_z v_y \rangle$ as follows:
\begin{eqnarray}
\langle v_x v_y \rangle &=&  \frac{\tau_f}{(2 \pi)^3 \mathcal{A}} \int d^3 {\bf k} \{V_{11} \phi_{11} + V_{12} \phi_{12} + V_{22} \phi_{22} \} \; , \\ \nonumber
\langle v_z v_y \rangle &=& \frac{\tau_f}{(2 \pi)^3 \mathcal{A}} \int d^3 {\bf k} \{H_{11} \phi_{11} + H_{12} \phi_{12} + H_{22} \phi_{22} \} \; ,
\end{eqnarray}
where the $\phi_{ij}$ functions are the power spectra associated with the forcing functions $h_1$ and $h_2$:
\begin{equation}
\langle \tilde{h}_i({\bf k_1},t_1) \tilde{h}_j({\bf k_2},t_2) \rangle = \tau_f \, (2\pi)^3 \delta({\bf k_1}+{\bf k_2}) \, \delta(t_1-t_2) \, \phi_{ij}({\bf k_2}) \; ,
\end{equation}
for $i$ and $j = 1,2$. The functions $V_{ij}$ and $H_{ij}$ (for $i$ and $j = 1, 2$) are expressed as integrals over the variable $\tau$, for example, $V_{12}$ can be written:
\begin{eqnarray}
\label{Decomp1}
\nonumber 
V_{12} &\equiv& \int_a^{+\infty} \; d\tau  e^{- 2 \nu \{Q(\tau)-Q(a)\}} \Bigl\{ \frac{(\epsilon^2-\gamma) \phi + \epsilon[\epsilon^2+1-\phi^2] \tau}{(\gamma+\epsilon^2)^2}  \\ 
&& \qquad - \frac{2 \epsilon \phi^2}{(\gamma+\epsilon^2)^{5/2}} \Bigl[T(\tau)-T(a)\Bigr] \Bigr\} \; ,
\end{eqnarray}
where, $a = k_x /k_y$, $\phi = (k_z - \epsilon k_x)/k_y$ and $\gamma=1+\phi^2$. Following \citet{Kim05}, this integral will be estimated in the strong shear limit, {\em i.e.} for $\xi \equiv \nu k_y^2 / \mathcal{A} \ll 1$. Thus, $\nu \{Q(\tau)-Q(a)\} = \xi [(\epsilon^2+1)\tau^3/3+\epsilon\phi\tau^2+\gamma \tau -\{\gamma a + (\epsilon^2+1)a^3/3  + b \epsilon a^2)\}]$ where $b=k_z/k_y$. Taking the limit $\xi \rightarrow 0$, we evaluate each term in (\ref{Decomp1}) by keeping only terms to leading order in $\xi$. The results are:
\begin{eqnarray}
\nonumber
V_{11} &=& - \frac{\gamma + \epsilon \phi a}{2 (\gamma + \epsilon^2)^2 R(a)} - \frac{\epsilon \phi}{2 (\gamma +\epsilon^2)^{5/2}} \kappa + \frac{(\gamma-\epsilon^2) \phi^2}{2 (\gamma+\epsilon^2)^3} \kappa^2  \\ \nonumber 
&& +  \frac{\epsilon \phi^3}{(\gamma+\epsilon^2)^3} \kappa^2 \mathcal{G} + \frac{\epsilon\phi(\phi^2-1-\epsilon^2)}{(\gamma+\epsilon^2)^{5/2}(\epsilon^2+1)} \kappa \mathcal{L} \; , \\ \nonumber
V_{12} &=& \frac{(\epsilon^2 - \gamma) \phi}{(\gamma+\epsilon^2)^{5/2}}\kappa - \frac{2 \epsilon \phi^2}{(\gamma + \epsilon^2)^{5/2}} \kappa \mathcal{G} + \frac{\epsilon(1+\epsilon^2-\phi^2)}{(\gamma+\epsilon^2)^2(\epsilon^2+1)} \mathcal{L}   \; , \\
\label{Trucszarb} 
V_{22} &=& \frac{\phi \epsilon}{(\gamma+\epsilon^2)^2} \mathcal{G} \; , \\ \nonumber
H_{11} &=&  \frac{\phi a - \epsilon}{2 (\gamma + \epsilon^2)^2 R(a)} + \frac{\phi}{2 (\gamma +\epsilon^2)^{5/2}} \kappa + \frac{\epsilon \phi^2}{(\gamma+\epsilon^2)^3} \kappa^2 \\ \nonumber 
&& - \frac{\phi^3}{(\gamma+\epsilon^2)^3} \kappa^2 \mathcal{G} + \frac{(\epsilon^2+1-\phi^2)\phi}{(\gamma+\epsilon^2)^{5/2} (\epsilon^2+1)} \kappa \mathcal{L} \; , \\ \nonumber
H_{12} &=& - \frac{2 \epsilon \phi}{(\gamma+\epsilon^2)^{5/2}}\kappa + \frac{2 \phi^2}{(\gamma + \epsilon^2)^{5/2}}  \kappa \mathcal{G}  + \frac{\phi^2-1-\epsilon^2}{(\gamma+\epsilon^2)^2(\epsilon^2+1)} \mathcal{L} \; , \\ \nonumber
H_{22} &=& - \frac{\phi}{(\gamma+\epsilon^2)^2} \mathcal{G} \; .
\end{eqnarray}
Here, $\kappa = \pi/2 -T(a)$. To derive Eq. (\ref{Trucszarb}), we used formula (2.173), (2.175) and (2.18) in \citet{GR65}. The results are written in terms of new integrals by making the substitution $y=2\xi(\epsilon^2+1)\tau^3/3$ and taking the limit $\xi \rightarrow 0$:
\begin{eqnarray}
\mathcal{G} &=&  \frac{1}{3} \Bigl(\frac{3}{2\xi(\epsilon^2+1)}\Bigr)^{1/3}  \int_0^\infty \frac{dy}{y^{2/3}} \times \\ \nonumber
&& \exp\Bigl[-y-\frac{3^{5/3} \epsilon \phi}{(\epsilon^2+1)} \Bigl(\frac{2\xi(\epsilon^2+1)}{3}\Bigr)^{1/3} y^{2/3}+O(\xi^{4/3})\Bigr]   \\ \nonumber
&=&  \frac{\Gamma(1/3)}{3} \Bigl(\frac{3}{2\xi(\epsilon^2+1)}\Bigr)^{1/3}  - \frac{3^{2/3} \epsilon \phi}{(\epsilon^2+1)} + O(\xi^{2/3}) \\ \nonumber 
&\equiv& \mathcal{G}_0 - \frac{3^{2/3} \epsilon \phi}{(\epsilon^2+1)} + O(\xi^{2/3}) \; , \\ \nonumber
\mathcal{L} &=& \frac{1}{3} \int_{\xi \rightarrow 0}^\infty \frac{dy}{y} \exp\Bigl[-y-\frac{3^{5/3} \epsilon \phi}{(\epsilon^2+1)} \Bigl(\frac{2\xi(\epsilon^2+1)}{3}\Bigr)^{1/3} y^{2/3}+O(\xi^{4/3})\Bigr]  \\ \nonumber
&=& \frac{-\ln(\xi)}{3} + O(\xi^{1/3} \ln(\xi))  \equiv \mathcal{L}_0 + O(\xi^{1/3}\ln(\xi)) \; .
\end{eqnarray}

We simplify (\ref{Trucszarb}) by keeping terms up to at most second order in $\epsilon$. To this order, we have:
\begin{eqnarray}
\phi &=& b-\epsilon a \; , \qquad  \gamma =1+b^2-2\epsilon ba  + \epsilon^2 a^2 \; , \\ \nonumber
&& \qquad R(a) = (\gamma+a^2)  \; , \\ \nonumber
\kappa &=& \Bigl[\pi/2 - \arctan\Bigl(\frac{a}{\sqrt{\gamma}}\Bigr)\Bigr] - \frac{\epsilon \phi \sqrt{\gamma}}{\gamma+a^2} + O(\epsilon^2) \equiv  \kappa_0 - \frac{\epsilon b}{\sqrt{g}} + O(\epsilon^2) \; . \\ \nonumber
\end{eqnarray}
For simplicity, we assume that $\phi_{ij}(k_y) = \phi_{ij}(-k_y)$ and consequently neglect all the terms proportional to an odd power of $b$ only (but keep those proportional to $b a$). Recalling that $\xi$ is a small parameter, we will keep only the terms proportional to $\xi^{-1/3}$ and $\xi^0$ and the leading order term in the $\epsilon$ expansion. The dominant contributions then give us:
\begin{eqnarray}
\label{TrucZarb2}
V_{11} &=& - \frac{1}{2 g(g + a^2)} +  \frac{b^2}{2 g^2} \kappa_0^2 - \frac{2 \epsilon^2 b^4}{g^{7/2}} \kappa_0 \mathcal{G}_0 + \dots \; , \\ \nonumber
V_{12} &=& - \epsilon \frac{2  b^2}{g^{5/2}} \kappa_0 \mathcal{G}_0 + \dots \; , \\ \nonumber
V_{22} &=& - \frac{3^{2/3} \epsilon^2 b^2}{g^2} + \dots \; , \\ \nonumber
H_{11} &=&  \frac{b a}{2 g^2 (g + a^2)} + \frac{2 \epsilon b^4}{g^{5/2} (g+a^2)} \kappa_0 \mathcal{G}_0  + \dots  \; , \\ \nonumber
H_{12} &=& \frac{2 b^2}{g^{5/2}}  \kappa_0 \mathcal{G}_0  + \dots \; , \\ \nonumber
H_{22} &=& \frac{3^{2/3} \epsilon b^2}{g^2} + \dots \; .
\end{eqnarray}
Here, $g=1+b^2=k_H^2/k_y^2$ and the dots stand for higher order terms in $\xi$ and $\epsilon$ compared to those which were kept.

We now express $\phi_{ij}$ in terms of $\psi_{ij}$ by assuming the forcing to be incompressible, for simplicity. In that case, the following relations hold: $\hat{h_1}(\tau) = R(\tau) \{\hat{f_x} + \epsilon \hat{f_z}\}$, and $\hat{h_2}(\tau) = [\phi \tau - \epsilon] \hat{f_x} + [\gamma+\epsilon\phi \tau] \hat{f_z}$. Keeping only the terms at most proportional to $\epsilon$, we obtain
\begin{eqnarray}
\label{Boforcing}
\phi_{11} &=& (g+a^2)^2 \psi_{11} + 2\epsilon (g+a^2)^2 \psi_{13} + O(\epsilon^2) \; , \\ \nonumber
\phi_{12} &=& (g+a^2) [b a \psi_{11} + g \psi_{13}] \\ \nonumber 
&& + \epsilon\{-(a^2+1)(g+a^2) \psi_{11}  + g (g+a^2) \psi_{33}\}  + O(\epsilon^2) \; , \\ \nonumber
\phi_{22} &=& b^2 a^2 \psi_{11} + 2 g b a \psi_{13} + g^2 \psi_{33} + \epsilon \{- 2 b a (a^2+1) \psi_{11} \\ \nonumber 
&&  - 2[a^2 b^2 + g(1+a^2)] \psi_{13} - 2 g b a \psi_{33}\}  + O(\epsilon^2) \; ,
\end{eqnarray}
By using (\ref{Boforcing}) and Eq. (\ref{TrucZarb2}) for the turbulent viscosities and by keeping only the dominant term for each component of the forcing, we obtain equations (\ref{EddyViscos}) given in the main text.

\bibliographystyle{aa}
\bibliography{../../../Biblio/Bib_sun,../../../Biblio/Bib_maths,../../../Biblio/Bib_dynamo,../../../Biblio/Bib_shear}

\end{document}